\documentclass[english,aps,prper,reprint,showpacs,longbibliography, floatfix]{revtex4-2}
\usepackage[utf8]{inputenc}
\usepackage{booktabs}
\usepackage{tabularx}
\usepackage{ragged2e}
\usepackage[T1]{fontenc}
\usepackage{geometry}
\usepackage{times}
\usepackage{hyperref}
\hypersetup{colorlinks=true,urlcolor=blue,citecolor=blue,linkcolor=blue}
\usepackage{array}
\usepackage{enumerate}
\usepackage{amsmath}
\usepackage{amssymb}
\usepackage{tikz}
\usepackage{graphicx}
\usepackage{multirow}
\usepackage{tcolorbox}
\usepackage{float}
\usepackage[normalem]{ulem}
\usepackage{setspace}
\usepackage{epstopdf}
\usepackage{mdframed}

\begin{document}

\title{A bottom-up taxonomy of student discourse with a Socratic AI physics tutor}

\author{Syed Furqan Abbas Hashmi}
\email{hashmi0@purdue.edu}
\affiliation{Department of Physics and Astronomy, Purdue University, West Lafayette, IN 47907, USA}

\author{N.~Sanjay Rebello}
\email{rebellos@purdue.edu}
\affiliation{Department of Physics and Astronomy, Purdue University, West Lafayette, IN 47907, USA}
\affiliation{Department of Curriculum and Instruction, Purdue University, West Lafayette, IN 47907, USA}

\begin{abstract}
Large language model (LLM) tutors are being deployed in introductory physics courses at a scale that produces transcript corpora far larger than traditional qualitative coding can absorb. A central question for physics education research (PER) is empirical and prior to any claim about effectiveness: \emph{what do students actually say to these tutors?} We address this question for one Socratic AI tutor deployed in an introductory calculus-based mechanics course by building a bottom-up taxonomy of student discourse. Each student turn is assigned an emergent free-text label by an LLM coder using the surrounding conversational context; near-paraphrase labels are then consolidated into a smaller set of discourse categories using a similarity-based grouping procedure. The procedure is validated against a stratified human-coded sample (Cohen's $\kappa = 0.78$). The resulting taxonomy of 357 categories is strikingly concentrated: the top 25 categories cover roughly half of all student turns, and two thematic bands: \emph{equation-handling} and \emph{meta-procedural requests} together dominate the head of the distribution. The substantive contribution is the taxonomy itself: a description of the discourse PER researchers can expect to encounter when students work with an AI tutor of this design, including a striking prevalence of meta-procedural turns in which students cede strategic control to the tutor.
\end{abstract}

\maketitle

\onecolumngrid
\clearpage
\twocolumngrid

% =====================================================================
\section{Introduction}
Large language model (LLM) tutors are no longer a thought experiment in physics education~\cite{kestin2025aitutoring, Letourneau2025}. Under controlled conditions, recent randomized trials find AI tutors producing learning gains that exceed in-class active learning~\cite{kestin2025aitutoring}, and beyond the lab, chatbots are already being studied in authentic introductory courses~\cite{kregear2025lab}. As these systems scale from pilots to course-wide deployments, they generate transcript corpora that dwarf what hand-coded qualitative analysis can absorb: a single semester of a single course routinely produces tens of thousands of student turns.

Physics education research (PER) has a mature theoretical vocabulary for student problem solving and discourse, expert/novice contrasts~\cite{larkin1980expert,chi1981categorization}, problem description schemas~\cite{heller1984prescribing}, symbolic forms~\cite{sherin2001equations}, epistemic games~\cite{tuminaro2007epistemic}, and epistemological framing~\cite{scherr2009framing}. However, its empirical machinery was built for transcripts of dozens of students, not thousands of sessions. Before PER can evaluate whether AI tutors help students learn, or test whether existing frameworks transfer to AI-mediated dialogue, there is a more basic empirical question to answer: \emph{What do students actually communicate to an AI physics tutor, and how frequently do they do it?}

This paper proposes such a pipeline and applies it to one corpus as a proof of concept. Specifically, we address two research questions:

\begin{itemize}
\item[\textbf{RQ1.}] What categories of discourse do students produce when interacting with a Socratic AI physics tutor, and how is their frequency distributed?
\item[\textbf{RQ2.}] What do the dominant categories reveal about how students frame their engagement with the scaffolding provided by the Socratic AI tutor?
\end{itemize}

% =====================================================================
\section{Background and related work}
% =====================================================================
Five decades of PER have established that expert and novice physicists differ in how they parse and frame problems. Larkin et al.~\cite{larkin1980expert} and Chi, Feltovich, and Glaser~\cite{chi1981categorization} showed that experts categorize problems by deep principle while novices key on surface features. Heller and Reif~\cite{heller1984prescribing} demonstrated that explicit problem description schemes substantially improve novice performance. Sherin's symbolic forms account~\cite{sherin2001equations} unpacked how students read structural meaning into equations, and Tuminaro and Redish~\cite{tuminaro2007epistemic} formalized recurring problem-solving patterns as ``epistemic games,'' including the ``Recursive Plug-and-Chug'' game in which students manipulate symbols with little reference to physical meaning. Scherr and Hammer~\cite{scherr2009framing} showed that students' epistemological framing of an activity, whether they treat it as algorithmic, conceptual, or social, is dynamically cued and can be tracked through observable behavior. Our pipeline targets precisely these constructs at scale.

Nelson's computational grounded theory framework~\cite{nelson2020cgt} formalized the iteration between unsupervised computational pattern-finding and human interpretation in qualitative social-science work. Tschisgale, Wulff, and Kubsch~\cite{tschisgale2023cgt} adapted this framework to PER, demonstrating that AI-assisted clustering can support, rather than replace, the interpretive work of qualitative coding. This is the most direct methodological precedent for the present paper. Odden, Marin, and Caballero~\cite{odden2020thematic} used topic models to map 18 years of PERC proceedings; Odden et al.~\cite{odden2024embeddings} extended this to deductive qualitative coding via text embeddings on student writing. Fussell, Stump, and Holmes~\cite{fussell2024trustworthy} provide a complementary contribution, showing how to assess the trustworthiness of machine coding through human--machine agreement diagnostics; we adopt their stance that automated coding requires explicit reliability claims rather than blind use, and report agreement statistics throughout the methods below.

Kortemeyer~\cite{kortemeyer2023ai} first showed that an off-the-shelf LLM could partially pass an introductory physics course, motivating their use as tutoring backends. Kestin et al.~\cite{kestin2025aitutoring} reported a randomized controlled trial in which an LLM tutor, designed around research-based principles, produced learning gains exceeding in-class active learning. Kregear, Babayeva, and Widenhorn~\cite{kregear2025lab} analyzed student--LLM interactions in an introductory lab, foregrounding question types and correctness. These studies motivate the deployment side; \textit{what is still missing is an analytic pipeline that converts the resulting transcripts into PER-interpretable structure}. This paper, building on our prior work, \cite{Hashmi2025} proposes a pipeline for interpreting student interactions with an AI Socratic tutor.

% =====================================================================
\section{Method}
% =====================================================================

\subsection{Setting and dataset}

% The corpus comes from a custom Socratic AI tutor built as a Django web application with a PostgreSQL backend and a retrieval-augmented generation pipeline over course materials, meaning the LLM's responses are conditioned on relevant excerpts from course material in addition to the student's question. The tutor was deployed as an optional study aid in an introductory calculus-based mechanics course (PHYS~172, Modern Mechanics) at a large midwestern public research university in Fall 2025. The system prompt instructed the model to use Socratic prompting and to scaffold conceptual setup before symbolic execution. Logging captured all turns at the message level. The corpus analyzed here consists of 5{,}513 total messages across 221 student sessions, of which 2{,}874 are student turns; the remaining messages are tutor responses. The analysis below operates on the 2{,}874 student turns.

The context of our study was an introductory calculus-based mechanics course for future engineers at a large U.S. midwestern public research university in Fall 2025. The course enrollment was 1,508 students. The course consisted of recitations in which students worked in groups of 3-5 students to solve a challenging Recitation problem, often an adaptation of a context rich problem \cite{bangs2012teaching}. A graduate teaching assistant (TA) walked around the room ready to answer questions.  However, given the low TA to student ratio, we deployed a Socratic AI tutor, as an optional study aid, to help students with the problem-solving process. For this study, students were saked to use the tutor to sove the REC as an extra credit activity. The AI tutor was built with Retrieval Augmented Generation and was based on our previous iteration described in~\cite{Hashmi2025}.

\begin{figure}
    %\begin{mdframed}
        \centering
        \includegraphics[width=\linewidth]{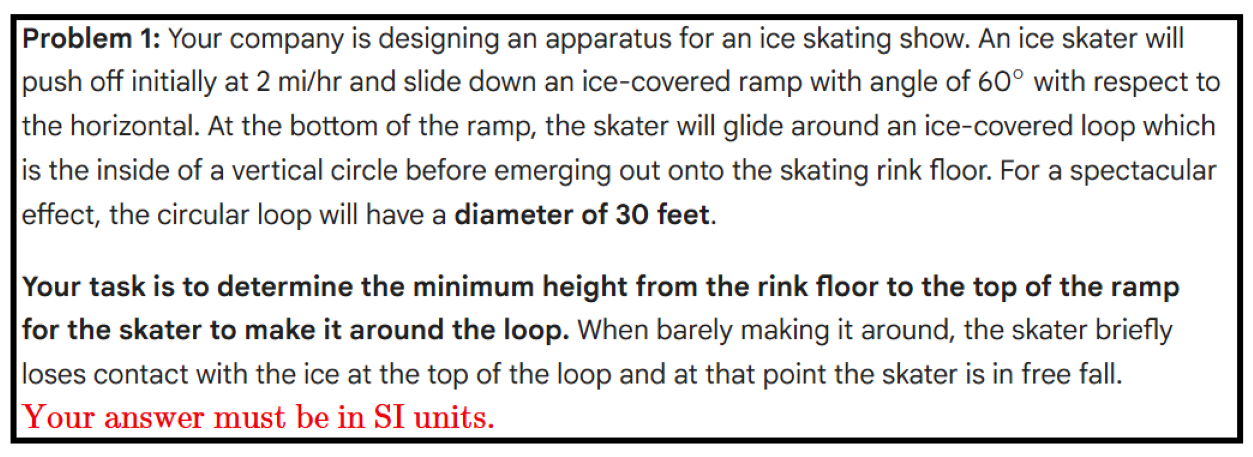}
        \caption{Recitation problem solved by students using the Socratic AI tutor}
    %\end{mdframed}
    \label{fig:Fig01}
\end{figure}

Our corpus of data comes from the Socratic AI tutor in the Recitation session of Week 8 of the 16-week semester.The Socratic AI tutor was deployed in five of the 34 Recitations sections of the course. Students used the tutor primarily to work through a multi-step problem assigned in the Recitation. The complete problem statement is shown in Fig ~\ref{fig:Fig01}. Solving the problem requires identifying the critical condition at the top of the loop (gravity supplies the centripetal acceleration), applying energy conservation between the top of the ramp and the top of the loop, and combining the two to solve for the minimum height. 

%an ice skater pushes off at 2~mi/hr at the top of a $60^\circ$ ice-covered ramp and slides down into a vertical circular loop of diameter 30~ft, and the student is asked to determine the minimum height of the ramp above the rink floor for the skater to make it around the loop, given that at the critical point the skater briefly loses contact with the ice at the top of the loop and is in free fall. 

Our corpus of data comes from a custom Socratic AI tutor built as a Django web application with a PostgreSQL backend and a retrieval-augmented generation pipeline over course materials, meaning the LLM's responses are conditioned on relevant excerpts from course material in addition to the questions asked by students. 

The system prompt instructed the model to use Socratic prompting and to scaffold conceptual setup before symbolic execution. Logging captured all turns at the message level. The corpus analyzed here consists of 5{,}513 total messages from 240 individual students across 221 completed student sessions, of which 2{,}874 are student turns; the remaining messages are tutor responses. The analysis below operates on the 2{,}874 student turns.

\subsection{Step~1: Emergent LLM discourse coding}

Rather than coding student turns against a predefined codebook, we asked an LLM (GPT-5.4-mini, accessed via the OpenAI API) to read each turn and produce its own short free-text label naming the discourse move the student was making, for example \emph{``proposing a force decomposition,''} \emph{``asking for confirmation,''} or \emph{``computing intermediate height.''} This is the LLM analog of open coding in grounded theory~\cite{nelson2020cgt,tschisgale2023cgt}: categories are allowed to emerge from the data rather than being fixed in advance.

A fixed codebook commits in advance to what is interesting about student discourse with an AI tutor, when the more honest first question is what is even in the data. The emergent approach lets the categories of student behavior be discovered rather than projected.

An isolated student turn is often ambiguous (\emph{``yes''} can be a confirmation, an agreement to proceed, or a restart cue). To stabilize labels against this single-message ambiguity, the prompt provided the previous four conversational turns (both student and tutor) as context. We piloted context windows of 0, 2, 4, and 8 turns on a held-out subset; four turns was the smallest window that recovered consistent labels for the most context-dependent moves without unnecessarily inflating prompt cost.

This step produced 833 unique raw labels across 2{,}874 messages. The category labels recovered by the taxonomy are shown in Table ~\ref{tab:top20}. Many of these labels (e.g., \emph{Writing Energy Equation}, \emph{Centripetal Relation Setup}, \emph{Solving for Height}, \emph{Threshold Speed Expression}) reflect this problem's structure. The label distribution was heavy-tailed by construction: many labels appeared only once or twice and were near-paraphrases of more common moves (\emph{``writing energy equation''}, \emph{``setting up energy equation''}, \emph{``writing energy conservation equation''}). The next step in the analysis process consolidates the labels.

\subsection{Step~2: Embedding-based label consolidation}

The 833 raw labels needed to be grouped so that semantically equivalent labels are treated as the same category. We did this in two stages: a similarity-based grouping pass, agglomerative clustering, followed by an LLM-generated name for each resulting cluster.

\subsubsection{Similarity-based Grouping}

Each raw label was converted into a \emph{text embedding}, which is a vector produced by a neural network (OpenAI's \texttt{text-embedding-3-small}) such that labels with similar meaning end up as similar lists of numbers, and unrelated labels end up as dissimilar vectors. The intuition is a map on which related phrases sit near each other and unrelated phrases sit far apart. The embedding gives each label its coordinates on that map. We computed the cosine distance, a standard similarity score based on the angle between two vectors, ranging from 0 (identical) to 1 (unrelated) between every pair of the 833 labels.

\subsubsection{Agglomerative Clustering}

We then applied \emph{agglomerative clustering}, a bottom-up grouping procedure that begins with every label as its own group and repeatedly merges the two closest groups until no remaining pair is closer than a chosen distance threshold. We used average-linkage merging, two groups are considered ``close'' if their members are close on average. We implemented this in \texttt{scikit-learn} with \texttt{distance\_threshold = 0.40}. Average-linkage was chosen because we expected groups of unequal size and shape (a few large discourse families, many small ones), and the distance-threshold formulation lets us pick how aggressively to merge without committing to a fixed group count.

The threshold was selected by sweeping over $[0.20, 0.60]$ in steps of 0.05. At each setting, the first author manually inspected a stratified sample of 50 groups and judged each as semantically coherent or not. The 0.40 threshold was selected as the largest setting at which $\geq 90\%$ of inspected groups were judged coherent (47 of 50); thresholds above this began merging mechanically related but discourse-distinct moves, while thresholds below it left near-paraphrases unmerged. We acknowledge that single-rater coherence judgments establish only face validity; a fuller study with multiple independent raters is left for future work.

\subsubsection{LLM-generated Nomenclature}

Each resulting group was then passed to an LLM call with the constituent raw labels and their counts, and the model was asked to produce a short canonical category name; cluster names were spot-checked by a human rater. This produced 357 consolidated discourse categories. The top 25 categories cover $\sim$52\% of all messages; categories with five or more messages cover 86\%. Long-tail categories with $\leq 4$ messages were retained in the codebook but are not the focus of the analyses below.

\subsection{Reliability against expert coding}

Following the trustworthiness recommendations of Fussell et al.~\cite{fussell2024trustworthy}, the LLM-generated labels were validated against a human expert who independently coded a stratified random sample of 287 student messages (10\% of the corpus, sampled to preserve session-level diversity and discourse-category balance). The expert produced free-text labels under the same prompt instructions given to the LLM. After both expert and LLM labels were mapped onto the consolidated category set produced above, agreement was Cohen's $\kappa = 0.78$ (raw agreement 84\%), which falls within the range conventionally interpreted as substantial agreement and is consistent with reliability levels reported for human and machine coding in recent PER work~\cite{tschisgale2023cgt,fussell2024trustworthy}. Disagreements clustered on borderline moves (e.g., confirmation vs.\ restatement) that we expected to be ambiguous in either direction.

% =====================================================================
\section{Findings}

The taxonomy itself is the principal empirical object of this paper. Table~\ref{tab:top20} lists the 20 highest-frequency consolidated categories, which together account for 47.2\% of all student turns. Two features of the head of the distribution warrant interpretation: its concentration, and its thematic composition.

Out of 357 consolidated categories the LLM identified in the data, the top 25 cover roughly half of all student turns, and categories with five or more messages cover 86\%. There is no prescribed dialogue path in the tutor's design, no canonical solution script, and no fixed sequence of prompts. Despite this, student discourse concentrates sharply on a small number of recurring moves. Across 221 independent sessions and the wide space of things a student \emph{could} say to a physics tutor, only a few dozen kinds of move account for most of what they actually said.

The top-20 categories sort into two thematic bands that together account for the bulk of the head of the distribution. We describe each in turn.

\begin{table}[t]
\caption{Top 20 consolidated discourse categories by message count. Percentages are of the 2{,}874 student turns. ``Raw'' refers to the number of distinct raw labels from Step 1 that merged into the category.}
\label{tab:top20}
\begin{ruledtabular}
\begin{tabular}{lrrr}
Category & $n$ & \% & Raw \\
\midrule
Writing Energy Equation         & 239 & 8.3\% & 31 \\
Next Step Guidance              & 127 & 4.4\% & 35 \\
Presenting Problem Statement    & 107 & 3.7\% & 12 \\
Velocity Solving                &  80 & 2.8\% & 49 \\
Solving for Height              &  68 & 2.4\% & 36 \\
Computing Heights               &  65 & 2.3\% & 42 \\
Tutor Interaction Checks        &  63 & 2.2\% & 34 \\
Restating Problem               &  61 & 2.1\% & 19 \\
Centripetal Relation Setup      &  50 & 1.7\% & 22 \\
Substitution and Simplification &  49 & 1.7\% & 23 \\
Requesting Solution Help        &  47 & 1.6\% & 27 \\
Collision Problem Solving       &  45 & 1.6\% & 31 \\
Algebra Problem Solving         &  44 & 1.5\% & 30 \\
Final Answer Reporting          &  43 & 1.5\% & 24 \\
Introducing New Problem         &  42 & 1.5\% & 14 \\
Asking for Principles           &  42 & 1.5\% & 20 \\
Energy Equation Setup         &  41 & 1.4\% & 18 \\
Threshold Speed Expression      &  39 & 1.4\% & 15 \\
Numerical Answering             &  38 & 1.3\% & 18 \\
Assumptions About Problem       &  37 & 1.3\% & 21 \\
\midrule
\textbf{Top-20 total}           & \textbf{1357} & \textbf{47.2\%} & 521 \\
\end{tabular}
\end{ruledtabular}
\end{table}

\subsection{Band 1: Equation-handling and symbolic execution}

The single largest band in the top 20 is equation-handling: \textit{Writing Energy Equation} (8.3\%), \textit{Velocity Solving} (2.8\%), \textit{Solving for Height} (2.4\%), \textit{Computing Heights} (2.3\%), \textit{Centripetal Relation Setup} (1.7\%), \textit{Substitution and Simplification} (1.7\%), \textit{Identifying Energy Conservation} (1.6\%), \textit{Algebra Problem Solving} (1.5\%), \textit{Threshold Speed Expression} (1.4\%), \textit{Numerical Answering} (1.3\%), together with \textit{Final Answer Reporting} (1.5\%). Roughly a third of all student turns in the top 20 are engaging in one of these tasks: writing down a specific physical relationship, manipulating it algebraically, substituting values, or reporting a numerical result.

These results are consistent with Sherin's account~\cite{sherin2001equations} of student engagement with equations as the central site of meaning-making in introductory mechanics, and with the symbolic-manipulation behavior central to Tuminaro and Redish's~\cite{tuminaro2007epistemic} epistemic games. What is striking is the prevalence of these moves \emph{relative to} conceptual or sense-making moves, given that the tutor's system prompt explicitly instructs it to scaffold conceptual setup before symbolic execution. The category structure suggests that in the dialogue students actually produce, symbolic work occupies the foreground.

\subsection{Band 2: Meta-procedural requests}

A second large band consists of turns in which the student does not advance the problem themselves but instead asks the tutor what to do: \textit{Next Step Guidance} (4.4\%, the second-largest category overall), \textit{Requesting Solution Help} (1.6\%),  \textit{Asking for Principles} (1.5\%), and \textit{Assumptions About Problem} (1.3\%). These are not equation-writing or computation; they are explicit requests for the tutor to supply the next strategic move, the relevant principle, or the modeling assumptions.

This band is the most striking feature of the taxonomy. The tutor is designed around Socratic prompting, which is precisely the pedagogical commitment to \emph{withhold} direct answers and direction in favor of questions that elicit student reasoning. Yet \textit{Next Step Guidance} alone is the second-most common discourse move in the entire corpus. Read in the framing terms of Scherr and Hammer~\cite{scherr2009framing}, students appear to frame the activity as one in which the tutor is expected to direct the procedural flow, with their role being to execute supplied steps. Whether this represents a failure of the Socratic design to cue the intended frame, or a stable equilibrium that students settle into despite the design, is not something the taxonomy alone can resolve, but the taxonomy makes the phenomenon visible and quantifies it.

\subsection{What is \textit{not} in the head of the distribution}

We note briefly, and without overclaiming, that the top 20 contains essentially \textit{no} categories corresponding to explicit conceptual reasoning, prediction, comparison to similar problems, or critical engagement with the tutor's suggestions. \textit{Tutor Interaction Checks} (2.2\%) is interactional rather than substantive (i.e. greetings, attention checks). The absence of conceptual-reasoning categories at the top of the distribution is suggestive but should be interpreted with care: it may reflect what students actually do, or it may reflect that conceptual moves are spread thinly across many low-count categories in the long tail rather than absent. A targeted analysis of the long tail is left for future work.

\section{Conclusion \& Implications}

We sought to answer two research questions. In \textbf{RQ 1.} we inquired about the categories of discourse and their frequency when students interact with a Socratic AI physics tutor. We found that student discourse concentrates sharply across a small number of recurring moves: of 357 consolidated categories recovered from 2{,}874 student turns, the top 25 account for roughly half of all turns. Despite no prescribed dialogue path in the tutor's design, students converge on a narrow set of moves across 221 independent sessions.

\par In \textbf{RQ 2.} we inquired about what the dominant categories reveal about how students frame their engagement with the scaffolding provided by the Socratic AI tutor. We found that the head of the distribution is dominated by two thematic bands: equation-handling and symbolic execution---consistent with Sherin's~\cite{sherin2001equations} account of equations as the central site of meaning-making and Tuminaro and Redish's~\cite{tuminaro2007epistemic} epistemic games---and meta-procedural requests in which students ask the tutor what to do next, which principle to apply, or what to assume. When considered in terms of the framing described by Scherr and Hammer~\cite{scherr2009framing}, students appear to frame the activity as one in which the tutor directs the procedural flow and their role is to execute supplied steps.

\par The most consequential observation for the design of Socratic AI tutors is the prominence of \textit{Next Step Guidance} and related meta-procedural categories: a tutor explicitly designed not to direct students nevertheless elicits a discourse in which directing is the second-most-requested service.

\section{Limitations \& Future Work}

Despite the findings reported above and their implications for the use of Socratic AI tutors, this study has several limitations. The corpus is single-site, single-course, single-tutor, and is dominated by a single mechanics problem context; the specific category labels (e.g., \emph{Centripetal Relation Setup}) reflect this topical scope and will not transfer to tutors covering other content. The analysis is observational; nothing here licenses causal claims about what the tutor's design contributes to the patterns we observe. Reliability rests on a 10\% expert-coded sample for Step 1 and a single-rater coherence audit for Step 2. A fuller human--machine agreement study in the style of Fussell et al.~\cite{fussell2024trustworthy}, with multiple independent coders across the full corpus, is the natural next step. Finally, the consolidated codebook is a snapshot of one model's groupings; rerunning the consolidation with a different embedding model could shift category boundaries.

Natural next steps include positional analysis of categories within sessions (where in a session each move occurs), cross-tabulation against student outcomes, and extension of the taxonomy across additional topical scopes. Each of these builds on the descriptive base established here.

\begin{acknowledgments}
This work is supported in part by U.S. National Science Foundation grants 2111138 and 2300645. Opinions expressed are those of the authors and not of the Foundation.
\end{acknowledgments}

\onecolumngrid
\clearpage
\twocolumngrid
\bibliographystyle{apsrev4-2}
\bibliography{references}
\end{document}